\documentclass[final]{svjour3}
\usepackage{graphicx}
\usepackage{rotating}
\usepackage{amssymb}
\usepackage{mathptmx}
\usepackage[numbers]{natbib}
\makeatletter
\journalname{Journal of Low Temperature Physics}
\bibpunct{}{}{,}{s}{}{,}

\begin{document}

\title{Recent Advances in Frequency-Multiplexed TES Readout: Vastly Reduced Parasitics and an Increase in Multiplexing Factor with sub-Kelvin SQUIDs}

\titlerunning{Recent Advances in Frequency-Multiplexed TES Readout}

\author{T.~de~Haan$^1$ \and 
        A.~Suzuki$^1$ \and
        S.~T.~P.~Boyd$^2$ \and
        R.~H.~Cantor$^3$ \and
        A.~Coerver$^1$ \and
        M.~A.~Dobbs$^4$ \and
        R.~Hennings-Yeomans$^1$ \and
        W.~L.~Holzapfel$^5$ \and
        A.~T.~Lee$^{1,5}$ \and
        G.~I.~Noble$^4$ \and
        G.~Smecher$^{6}$ \and
        J.~Zhou$^1$}

\authorrunning{T.~de~Haan et al.}

\institute{$^1$ Physics Division, Lawrence Berkeley National Lab, CA, 94720, USA \\
           $^2$ Department of Physics \& Astronomy, University of New Mexico, Albuquerque, NM, 87131, USA \\
           $^3$ STAR Cryoelectronics, Santa Fe, NM, 87508, USA \\
           $^4$ Department of Physics and McGill Space Institute, McGill University, 3600 Rue University, Montreal,
QC H3A 2T8, Canada \\
           $^5$ Department of Physics, University of California, Berkeley, CA, 94720, USA \\
           $^6$ Three-Speed Logic, Inc., Vancouver, BC V6A 2J8, Canada \\
\email{tijmen@lbl.gov}}

\maketitle

\begin{abstract}

Cosmic microwave background (CMB) measurements are fundamentally limited by photon statistics. Therefore, ground-based CMB observatories have been increasing the number of detectors that are simultaneously observing the sky. Thanks to the advent of monolithically fabricated transition edge sensor (TES) arrays, the number of on-sky detectors has been increasing exponentially for over a decade. The next-generation experiment CMB-S4 will increase this detector count by more than an order of magnitude from the current state-of-the-art to ~500,000.
The readout of such a huge number of exquisitely precise sub-Kelvin sensors is feasible using an existing technology: frequency-domain multiplexing (fMux). To further optimize this system and reduce complexity and cost, we have recently made significant advances including the elimination of 4~K electronics, a massive decrease of parasitic in-series impedances, and a significant increase in multiplexing factor. %We discuss the remaining challenges and prospects for the future.
\keywords{transition edge sensors, frequency-domain multiplexing, readout electronics}

\end{abstract}

\section{Motivation}

Transition edge sensors (TES) are a mature and extremely successful technology for calorimetric and bolometric measurements. Their low operating temperature ($0.01-0.5\,\mathrm{K}$), low ($0.01-2\,\mathrm{\Omega}$) impedance, and stringent noise requirement ($\lesssim10\,\mathrm{pA/\sqrt{Hz}}$) calls for a sophisticated multi-stage readout system. As the per-experiment number of detectors starts to approach $\mathcal{O}(10^6)$,\cite{cmbs4} a highly multiplexed, simple to assemble, and cost-effective solution is required. 

Digital frequency-domain multiplexing \cite{SRON, dobbs11} uses an LC resonator in series with each TES detector to select a frequency at which a voltage bias is applied. This voltage bias keeps the detector under strong negative electrothermal feedback while also serving as a probe tone: the resulting current represents the signal of interest and is sensed with a SQUID amplifier. 

The stiffness of the voltage bias affects detector linearity, stability, and crosstalk. It is therefore important to minimize the Th\'evenin equivalent series impedance to the TES. This firstly requires minimizing any loss/resistance in the readout system other than the TES itself by making use of superconducting circuit elements, minimizing capacitor dielectric loss, and avoiding magnetic coupling to surrounding normal conductors. Secondly, effective real series impedance can result from purely reactive circuit elements. Inductances present in any individual leg of the multiplexing comb can be tuned out by a judicious choice of bias frequency, but inductances that are common to the comb can only partially be tuned out, leaving an effective real impedance in series with the TES. These inductances are the limiting factor in the achieved stability, multiplexing factor, linearity, and crosstalk of currently deployed fMux systems.\cite{pb2, spt3g_readout}

The focus of this work is to outline recently made advances in fMux readout technology, primarily dissipationless bias and sub-Kelvin SQUID operation.\cite{lowitz} These come with a host of advantages with the primary benefit being reduced parasitic impedance. Secondary benefits include eliminating all electronics on intermediate-temperature stages, ease of assembly, and eliminating the need for a low-inductance, low thermal conductivity cryogenic cable. The reduction of the parasitic impedance in the multiplexing circuit allows for higher multiplexing factors and/or lower impedance detectors.

In \S\ref{sec:cimm}, we present the test module we made for the purpose of demonstrating these technologies. \S\ref{sec:results} shows the results we obtained from this test module. We summarize these results and provide an outlook to the future in \S\ref{sec:conclusions}.

\begin{figure}
\begin{center}
\includegraphics[width=0.85\linewidth,keepaspectratio]{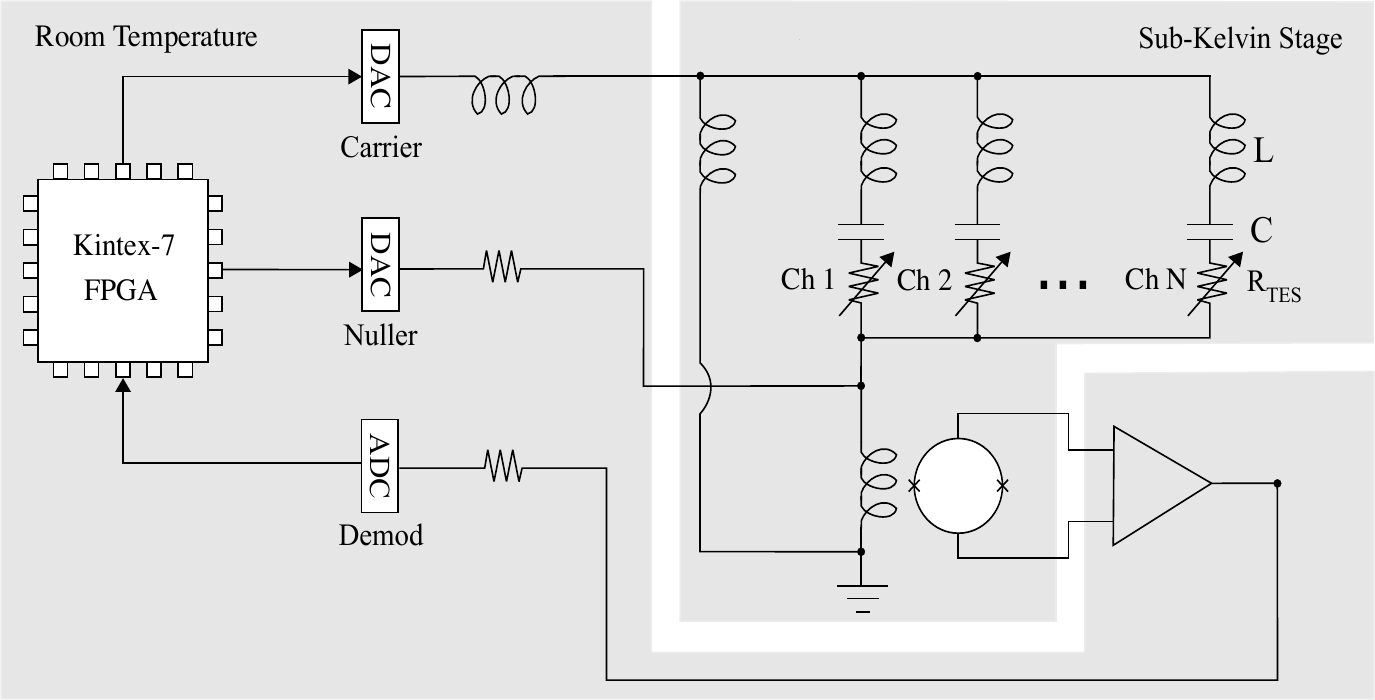}
\end{center}
\caption{Simplified circuit diagram adapted from \cite{dehaan12}. Note that the TES voltage bias is provided by a dissipationless inductive divider, and the low power dissipation, low input inductance SQUID amplifier is mounted on the same temperature stage as the TES detectors. By housing the SQUID and bias circuit on the sub-Kelvin stage, we drastically reduce the series impedance to the TES and eliminate the need for low-inductance cables.}
\label{fig:circuit_diagram}
\end{figure}

\section{Implementation}
\label{sec:cimm}

\begin{figure}
\begin{center}
\includegraphics[width=0.55\linewidth,angle=90]{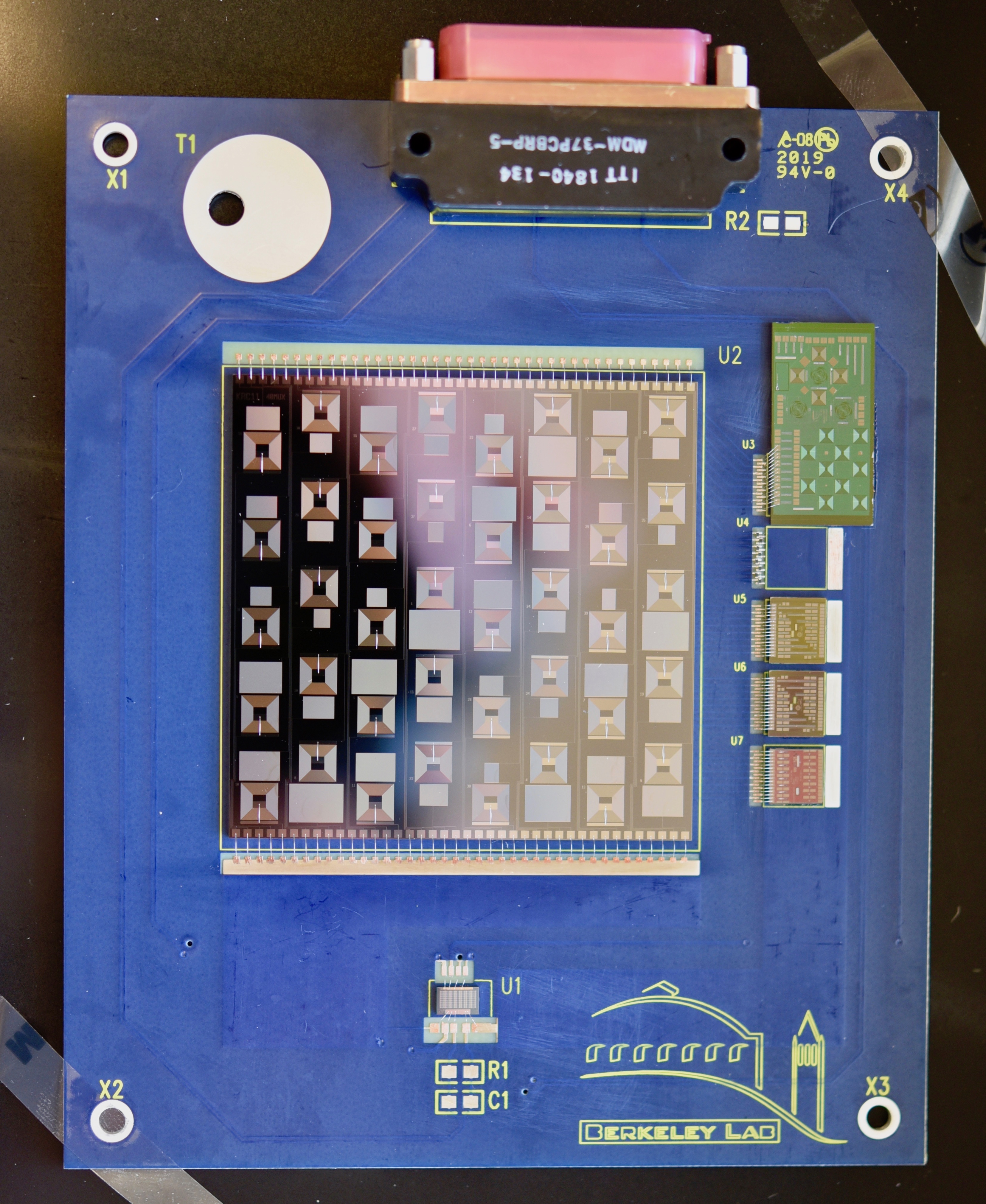}
\end{center}
\caption{Cold Integrated fMux Module (CIMM) after component assembly but prior to mounting into its enclosure. This compact assembly allows for end-to-end testing of the readout system with TES detectors in a light-tight, magnetically shielded environment. (Color figure online)}
\label{fig:cimm}
\end{figure}

We have implemented a Cold Integrated fMux Module (CIMM) for technology demonstration purposes. The CIMM hosts a full readout chain on the sub-Kelvin cryogenic stage, including a SQUID amplifier fabricated at STAR Cryoelectronics, LC resonators fabricated at the LBNL Microsystems Laboratory, and a variety of compatible TES detectors. These components are mounted on a PCB which provides superconducting signal routing, mechanical placement, and thermal anchoring. The signals are carried off the PCB using a micro-D connector and the board houses a thermometer and heater for thermal control. The assembled PCB is then enclosed in a superconducting, light-tight shield, which in turn is enclosed in a mu-metal magnetic shield. This completed package is then mounted on the sub-Kelvin cryogenic stage. Figure \ref{fig:cimm} shows the CIMM PCB after component mounting and wirebonding, prior to mounting into its enclosure. 

\section{Results}
\label{sec:results}

%The initial version of the CIMM has a multiplexing factor of 40x, and yields the following results.
% Upon testing the first version of the CIMM, which had a multiplexing factor of 40, we found promising results.

\subsection{SQUID performance}

The Series SQUID Array Amplifier (SSAA) shows the expected input coupling of approximately $25\,\mathrm{\mu A}/\Phi_0$ and an excellent forward gain of up to $1500\,\mathrm{V/A}$ depending on the choice of bias point. We determine that mu-metal shielding is important to achieving this level of performance. The power consumption is measured to be $\sim 20\,\mathrm{nW}$.

\subsection{Network Analysis}

The detector and LC resonator yield is 100\% with all channels showing a superconducting transition. The network analysis is fit by minimizing the $\chi^2$ between the measurement data and an analytic circuit model that includes a parameter for each resonator's individual capacitance and resistance, as well as five global parameters. This 85-dimensional fit gives acceptable agreement and shows that:
\begin{itemize}
    \item The stray impedance in series with the LCR comb is roughly $10\,\mathrm{nH}$.
    \item There is no measureable loss (real impedance) in series with the LCR comb.
    \item The capacitance across the LCR comb is roughly 200~pF.
    \item The bias inductor value is roughly 4.2~nH, compared to the 5~nH design target. 
\end{itemize}
Note that no quantitative uncertainties are presented as the formal statistical uncertainties are much smaller than the systematic uncertainties from using an imperfect analytic circuit model, the effects of which do not study in this work. 

\begin{figure}
\includegraphics[width=0.5\linewidth]{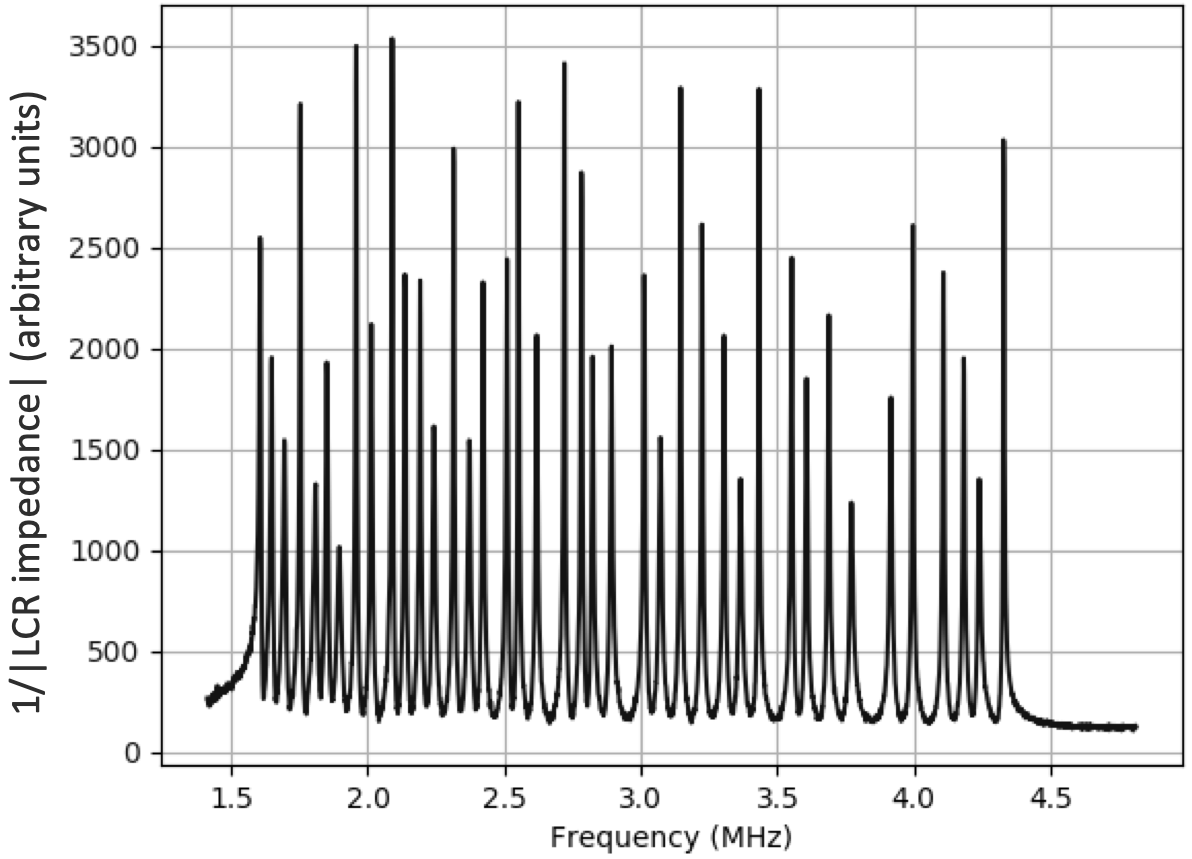}
\includegraphics[width=0.5\linewidth]{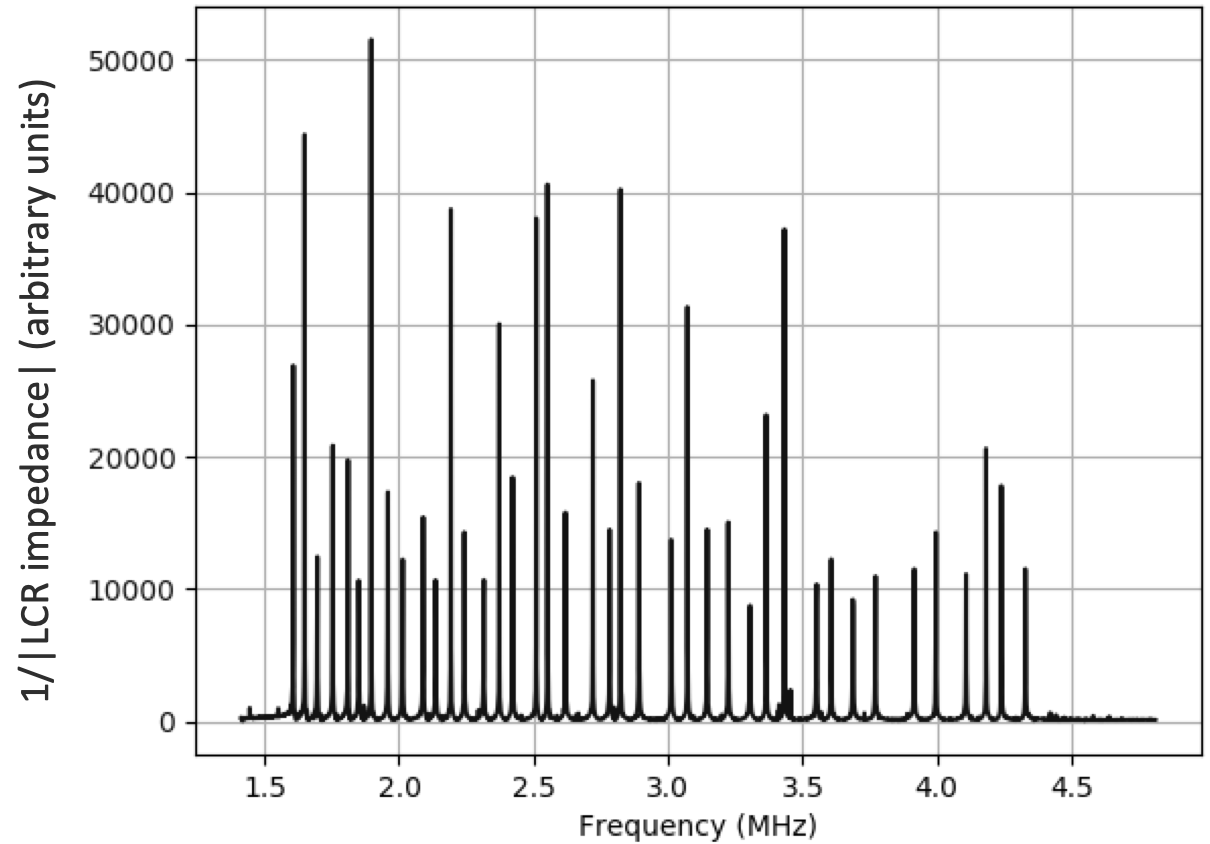}
\caption{Network analysis of the 40x CIMM. \emph{Left:} Measurement at 800~mK. The peak height non-uniformity is due to the varying intrinsic detector properties. \emph{Right:} Measurement at 250~mK. All peaks show a decrease in width and increase in height indicating the TES has undergone a superconducting transition. In this case, the peak height non-uniformity is due to the narrower peak widths falling below the frequency spacing of the measurement.}
\label{fig:test}
\end{figure}

\subsection{Detector Operation}

\begin{figure}
\begin{center}
\includegraphics[width=0.58\linewidth,height=0.42\linewidth]{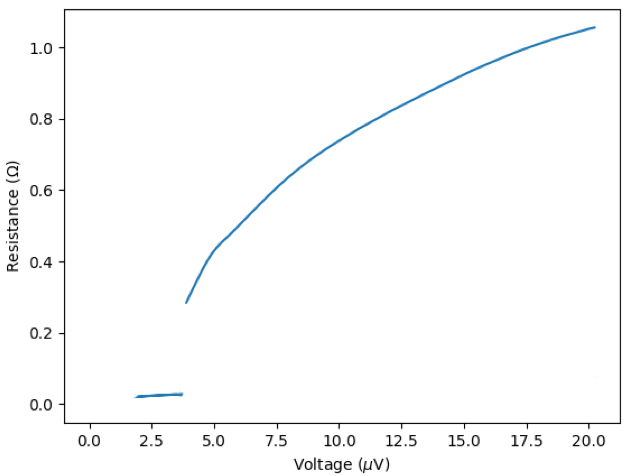}
\end{center}
\caption{Resistance-voltage curve showing a detector transition in the CIMM configuration. Compared to the previous generation of this readout system, the detector is able to drop more deeply into the superconducting transition prior to latching to a superconducting state, at a resistance of about 30\% of the normal resistance. When the TES is superconducting the residual measured impedance is approximately $0.02\,\mathrm{\Omega}$, representing an order of magnitude improvement.}
\label{fig:rv}
\end{figure}

Demonstration of TES detector operation in a loading-free environment such as the CIMM consists of warming the sub-Kelvin stage to above the critical temperature of the TES, applying a large voltage bias, lowering the stage temperature, then slowly decreasing the voltage bias until the detector resistance starts to fall. Normally, the voltage would be halted at a point where the detector has high responsivity and is sufficiently fast, but in this case we continue to drop the voltage in order to determine the point at which the detector latches into superconductivity. When the TES is superconducting, we measure the residual impedance of the circuit. Figure \ref{fig:rv} shows one such transition. In this case, the detector latches at around 30\% of its normal resistance, and the residual measured impedance is very low at $\sim 0.02\,\mathrm{\Omega}$. 
%This represents an order-of-magnitude improvement over currently deployed systems.

\subsection{Noise Performance}

\begin{figure}
\begin{center}
\includegraphics[width=0.6\linewidth,keepaspectratio]{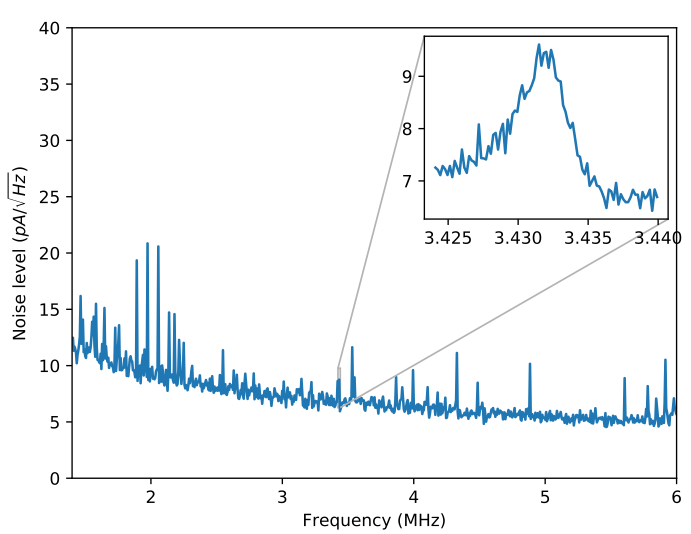}
\end{center}
\caption{Spectral density of the local white noise level. Note that the main figure does not have the spectral resolution to resolve the LC resonant peaks. The inset shows a high-resolution measurement of one of the resonant peaks, demonstrating the expected on-resonance increase of the noise level.}
\label{fig:noise}
\end{figure}

A key performance parameter of the fMux readout system is its noise equivalent current. This can be measured by, at each frequency, measuring the current spectral density at the output of the SQUID amplifier, then dividing by the effective forward gain of the SQUID. We test the CIMM in an RF-tight room and measure the noise as a function of frequency referred back to the TES detector, shown in Figure~\ref{fig:noise}. The white noise level is seen to be below the requirement of $12\,\mathrm{pA/\sqrt{Hz}}$. The noise in this case falls with frequency, but this is due to our choice of transfer function of the bias circuit, which also causes the available dynamic range to fall with frequency. We expect to correct this frequency dependence to create a dynamic range independent of bias frequency, which we expect to raise the noise level to approximately $10\,\mathrm{pA/\sqrt{Hz}}$ across the full readout bandwidth.

We find a slight excess in white noise level near the LC resonances. This is expected due to the LC resonators modulating the DAC noise, amplifier noise, and Johnson noise of the warm amplifier chains that produce the carrier and nuller.

\subsection{132x LC Resonators}

\begin{figure}
\begin{center}
\includegraphics[width=\linewidth, keepaspectratio]{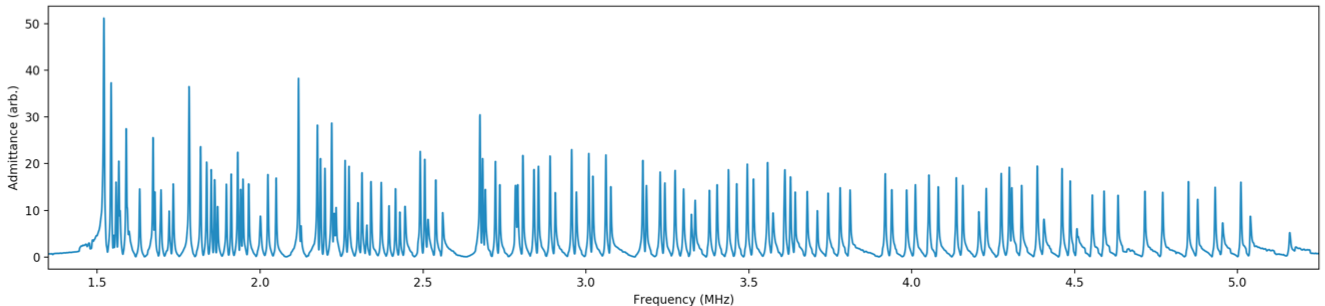}
\end{center}
\caption{Network analysis of a prototype resonator chip. Out of 130 possible resonances, 122 resonant peaks were observed, representing a 94\% yield. The non-uniformity of the peak spacing is expected to improve in the next iteration of the LC chip.}
\label{fig:netanal_132}
\end{figure}

Currently deployed iterations of the fMux readout system use multiplexing factors of up to 68 channels per SQUID element (68x). By increasing the multiplexing factor to 132x, we further reduce the cost and complexity of the readout system. Having demonstrated sub-Kelvin SQUID operation and inductive biasing, we have drastically lowered the parasitic impedance in series with the LCR comb, enabling a 132x test. We perform this test by placing an LC resonator chip in a 132x-appropriate CIMM with a SQUID amplifier, but with superconducting shorts in place of TES chips. The resulting network analysis is shown in Figure \ref{fig:netanal_132}.
Out of the 132 designed resonances, two channels were not connected, and 122 resonant peaks were observed, representing a lower limit on the LC resonator yield of 94\%.

\section{Conclusions and Outlook}
\label{sec:conclusions}

We have implemented an improved frequency-domain multiplexing readout scheme that uses reactive biasing and sub-Kelvin SQUID amplification. This new scheme reduces the critical performance parameter, the parasitic inductance in series with the LCR comb, by an order of magnitude down to $\sim10\,\mathrm{nH}$. This reduction in parasitic inductance improves detector linearity, stability, and crosstalk, and enables advances such as an increased multiplexing factor. The difficult low thermal-conductance and low-inductance cables between the sub-Kelvin and 4~Kelvin stages are no longer required, simplifying the ease of implementation and assembly of the readout system. We have shown good detector performance by meeting the noise requirement and demonstrating high yield.

% optical CIMM
% low impedance detectors
% 132x v2

In the near future, we plan to use the CIMM to test already-fabricated, lower-impedance detectors with normal resistance values around $200\,\mathrm{m\Omega}$. The operation of these devices is enabled by the low-parasitic system demonstrated in this work. We expect that these lower-impedance detectors will make the system more robust to any potential excess current noise in the system, as the lower detector impedance requires a lower voltage bias, raising TES responsivity.

The Simons Array project will include an on-sky demonstration of fMUX readout with sub-Kelvin SQUIDs and reactive biasing. This will lean heavily on the CIMM in its design, and is expected to be deployed in early 2020 on the Huan Tran Telescope reading out a small fraction of the POLARBEAR-2c experiment focal plane.

\begin{acknowledgements}
TdH acknowledges the LBNL Chamberlain Fellowship with funding from 
%the Quantum Sensors HEP-QIS Consortium 
Quantum Information Science Enabled Discovery (QuantISED) for High Energy Physics (KA2401032)
and the LBNL Laboratory Directed Research and Development (LDRD) program under U.S. Department of Energy Contract No. DE-AC02-05CH11231.
\end{acknowledgements}

% \pagebreak

\end{document}